\documentclass[conference]{IEEEtran}
\IEEEoverridecommandlockouts
\usepackage{placeins}
\usepackage{float}
\usepackage{cite}
\usepackage{amsmath,amssymb,amsfonts}
\usepackage{algorithmic}
\usepackage{graphicx}
\usepackage{placeins}
\usepackage{textcomp}
\usepackage{float}
\usepackage{doi}
\usepackage{hyperref}
\DeclareUnicodeCharacter{2082}{\ensuremath{_2}}
\def\BibTeX{{\rm B\kern-.05em{\sc i\kern-.025em b}\kern-.08em
    T\kern-.1667em\lower.7ex\hbox{E}\kern-.125emX}}
\begin{document}

\title{Energy and Performance Benchmarking of Deep Learning Models for Breast Cancer Detection\\
\thanks{Financial support was provided by the Mitacs
Globalink Research Internship Program and the Canadian Defence Academy Research Program.}
}

\author{\IEEEauthorblockN{1\textsuperscript{st} Samar Garrab}
\IEEEauthorblockA{\textit{Royal Military College of Canada} \\
\textit{Kingston, ON, Canada}\\
samar.garrab@rmc.ca}
\and
\IEEEauthorblockN{2\textsuperscript{nd} Ghada Achour}
\IEEEauthorblockA{\textit{National Engineering School of Sousse} \\
\textit{Sousse, Tunisia}\\
achour.ghada@eniso.u-sousse.tn}
}
\maketitle

\begingroup \renewcommand\thefootnote{} \footnotetext{\textbf{Accepted Manuscript Notice}\\
This is the accepted version of the paper accepted for publication in the 2026 IEEE International Conference on Machine Learning and Applications (ICMLA 2026). Copyright © IEEE. Personal use of this material is permitted. Permission from IEEE must be obtained for all other uses.} \endgroup

\begin{abstract}
Recent advances in machine learning have greatly improved breast cancer detection, enabling more accurate and timely diagnosis. Deep learning (DL) models show strong potential for medical image analysis; however, as their architectural complexity increases, their environmental impacts are becoming a growing concern.
In this paper, we present a comparative analysis of seven DL models for breast cancer detection on two medical datasets: Breast Ultrasound and BreakHis 400X. The evaluated architectures range from Convolutional Neural Networks (CNNs) and transformers to hybrid models. In addition to performance metrics, we assess CO$_2$ emissions during both training and inference.
Our results show that EfficientNet and ResNet consistently deliver strong performance, although with higher CO$_2$ emissions. The selected transformers, such as DeiT-Tiny, perform competitively on both datasets, whereas DenseNet121 achieves lower accuracy. On the Breast Ultrasound Dataset, DeiT provides the most favourable balance between accuracy and energy consumption, whereas on the BreakHis dataset, the ViT and Swin models achieve the best results.
Overall, our findings indicate that no single architecture category from the evaluated ones consistently dominates across the two selected datasets. Our results highlight the importance of jointly considering performance, emissions, and dataset characteristics when selecting models for medical applications.
\end{abstract}

\section{Introduction}
Artificial Intelligence (AI) and Deep Learning (DL), in particular, have emerged as powerful tools for solving complex pattern recognition problems, particularly in medical image analysis and segmentation~\cite{litjens2017survey,shin2016deep}. DL models have achieved remarkable success in healthcare applications, including disease diagnosis and medical image segmentation. Architectures such as Convolutional Neural Networks (CNNs) and transformers have become well-suited for this task, learning meaningful patterns directly from raw images~\cite{esteva2017dermatologist}.
Early work focused primarily on CNNs, but transformers have increasingly challenged their dominance. Gheflati and Rivaz~\cite{gheflati2022vit} were among the first to explore this shift in ultrasound data, showing that transformer models can match or exceed CNN performance in the Breast Ultrasound dataset.

Although DL models can provide excellent predictive performance, they usually incur considerable computational cost. Consequently, their energy consumption has become an important issue~\cite{strubell2019energy, schwartz2020greenai}.
This has led to Green AI, which emphasizes energy consumption, computational cost, and carbon emissions rather than focusing solely on predictive performance~\cite{schwartz2020greenai}. Beyond accuracy, Green AI asks how much energy a model consumes and at what environmental cost it operates. In healthcare, where AI systems often run in resource-constrained settings, these considerations are very important. 
Most existing research on breast cancer detection focuses on improving accuracy, paying little to no attention to energy and computational costs. Although sophisticated models often perform better, there is a clear trade-off between accuracy and efficiency, as they require more resources for training and inference~\cite{touvron2021deit, tan2019efficientnet}.
Furthermore, much research uses a single dataset or compares only a small number of models, limiting the breadth to which their findings can be applied. What the field needs is a broader comparison that examines multiple DL models and datasets and asks not just how well they perform but also at what cost.

The main contributions of this paper are as follows:
\begin{itemize}
\item We present a comprehensive comparison of seven DL models, including CNNs, transformers, and hybrid architectures, using two medical datasets with different imaging modalities: the Breast Ultrasound dataset and the BreakHis 400X dataset.
\item We evaluate both predictive performance (accuracy, F1-score, recall, and ROC) and energy consumption.
\item We highlight the trade-offs between performance and computational efficiency across the evaluated models and datasets to support the sustainable use of AI systems in healthcare.
\end{itemize}
The remainder of this paper is organized as follows. Section~\ref{sec:Methodology} presents the research methodology. Section~\ref{sec:ExperimentalSetup} reports the experimental setup, and Section~\ref{sec:ResultsDiscussion} presents and discusses the main results, highlighting the trade-offs between predictive performance and energy consumption. Finally, Section~\ref{sec:Conclusion} summarizes key findings and outlines directions for future research.

\section{Methodology}
\label{sec:Methodology}
This section presents the research methodology used in this paper, including datasets, DL models and evaluation metrics.
\subsection{Datasets}

Two publicly accessible breast cancer imaging datasets were used in this study. The Breast Ultrasound dataset~\cite{ultrasound-paper} comprises 780 ultrasound images in PNG format, each annotated into one of three categories (normal, benign, and malignant), and is therefore suitable for supervised breast lesion classification. The BreakHis 400X dataset~\cite{spanhol2016dataset} contains 1,700 high-resolution histopathological images acquired at 400$\times$ optical magnification and labeled as benign or malignant. Together, these datasets encompass complementary imaging modalities, enabling a comprehensive assessment of model performance on both ultrasound and histopathological breast cancer images.

\subsection{Deep Learning Architectures} 
This study assesses seven widely adopted DL architectures: three CNNs (ResNet, DenseNet, and EfficientNet), three transformers (ViT, DeiT, and Swin), and one hybrid architecture (ConvNeXt). Collectively, these models embody distinct architectural paradigms, ranging from purely convolutional designs to self-attention–based transformer frameworks and hybrid CNN–transformer configurations. This selection enables a comprehensive comparative analysis of contemporary architectures with respect to classification accuracy and computational efficiency in the context of breast cancer image analysis.

\subsection{Evaluation Methodology and Metrics}
As an evaluation methodology, this study assesses both predictive performance and computational efficiency for each selected model using both datasets.

\subsubsection{Energy Measurement}

To quantify the environmental impact associated with both training and inference, we used the CodeCarbon library~\cite{lacoste2019quantifying,garrab2026}. Carbon emissions were measured using the \textit{EmissionsTracker} API. The tracker was started immediately before the training or inference process and stopped upon its completion. During execution, CodeCarbon estimates the energy consumed by the computing hardware (CPU/GPU) and converts it to equivalent CO$_2$ emissions (CO$_2$eq) using the electricity carbon intensity of the execution environment. The following energy metrics were separately evaluated:
\begin{itemize}
    \item Training carbon emissions in grams CO$_2$ equivalent (g CO$_2$)
    \item Inference carbon emissions in g CO$_2$
\end{itemize}

\subsubsection{Accuracy Measurement}
The performance of each model is evaluated using standard metrics commonly used in medical image analysis.
\begin{itemize}
    \item Accuracy: The best validation accuracy was chosen.
    \item Weighted Precision: Proportion of predicted positive samples, truly positive, computed with class-frequency weighting.
     \item Weighted Recall: Measures how well the model identifies true positives across all classes.
     \item Weighted F1-score: Balances precision and recall into a single metric, weighting each class by its frequency.
     \item Area Under the ROC Curve (AUC): Measures how well the model separates between benign and malignant cases.
\end{itemize}
\section{Experimental Setup}
\label{sec:ExperimentalSetup}
This section describes the experimental setup, including dataset preparation, augmentation, implementation, and hardware.
\subsection{Data Preparation and Augmentation}
In this section, we will detail the data preprocessing, augmentation and normalization for each dataset.

\subsubsection{Breast Ultrasound Dataset}
To prepare the dataset, mask files were excluded, as they are not required for the image classification task. The remaining data were partitioned into training (70\%), validation (15\%), and test (15\%) subsets. All images were converted to PyTorch tensors and normalized using the ImageNet channel-wise mean and standard deviation. Data augmentation was applied exclusively to the training set to enhance model generalization, whereas the validation and test sets underwent only resizing and normalization.
To mitigate class imbalance, class weights were computed using the balanced option in scikit-learn and integrated into the loss function, thereby assigning higher penalties to minority classes, specifically the malignant and normal categories. The data augmentation pipeline include the following transformations: random horizontal flip (probability p = 0.5), random vertical flip (p = 0.2), random rotation (±15°), random affine transformation with translation (10\%) and scaling (0.9–1.1), color jitter with a brightness, contrast, and saturation factor of 0.2, and random conversion to grayscale (p = 0.1). All experiments were conducted using a fixed random seed (42) for Python, NumPy, and PyTorch to promote deterministic and reproducible behaviour.

\subsubsection{BreakHis 400X Dataset}
This dataset contains breast histopathological images classified as benign or malignant. Images from the original training and testing folders were merged into a single dataset, labeled with their folders, and cleaned by removing missing or unreadable files, resulting in 1,693 valid images. All experiments were conducted in Python using NumPy and TensorFlow, with a fixed random seed (42) and deterministic TensorFlow operations enabled to ensure reproducibility. The data augmentation strategy includes random horizontal and vertical flips, random rotation (0.15), random zoom (0.15), random contrast (0.15), and random brightness (0.10).
To ensure a robust and unbiased evaluation, a unique patient identifier was extracted from each file name by combining the second and third hyphen-separated fields. This enabled a patient-level split, preventing images from the same patient from appearing in multiple subsets and eliminating data leakage. Among the 81 patients, 57 (70\%) were assigned to the training set, and 12 (15\%) each to the validation and test sets. Data augmentation was applied only to the training set, while all images were resized and normalized according to each model's input requirements. In addition, training samples were reshuffled at every epoch to improve robustness despite the limited number of patients.

\subsection{Model Training and Implementation Details}
In this subsection, we detail the training configurations of both datasets and outline the hardware setup used.

\paragraph{Training Configuration for Breast Ultrasound Dataset}
Since the dataset is imbalanced, we used balanced class weights and passed them directly to the CrossEntropyLoss function, allowing the model to learn more effectively from underrepresented classes. For optimization, AdamW was used with a learning rate of 1e-4 for CNNs and of 5e-6 for transformers and hybrid architectures, which are known to be more sensitive to this setting and tend to train more stably~\cite{vit-lr}. A ReduceLROnPlateau scheduler was used to reduce the learning rate when validation loss plateaued. Training was performed for up to 100 epochs with a batch size of 8, shuffled data, and early stopping with a patience of 7 epochs to prevent overfitting.

\paragraph{Training Configuration for BreakHis Dataset}
To reduce overfitting and preserve pretrained visual representations, the backbone was kept frozen during training, which is important given the dataset's limited size and variability~\cite {tan2018survey}.
Since only the classification head is trained, optimization is more stable and efficient. Training uses a batch size of 32 and the AdamW optimizer, with learning rates of $10^{-4}$ for CNNs and $5\times10^{-4}$ for transformer and hybrid models. The model is trained for up to 100 epochs with EarlyStopping (patience of 7) based on validation AUC. A fixed seed of 42 was used for reproducibility. All models were implemented in Keras, except the Swin transformer and ConvNeXt, which were implemented in PyTorch.
\subsection{Hardware Setup and Computational Environment}
All experiments were conducted on the Digital Research Alliance of Canada Fir cluster using an NVIDIA H100 80GB HBM3 GPU, providing efficient training and evaluation of DL models.
\section{Results and discussion}
\label{sec:ResultsDiscussion}
This section presents and discusses the predictive performance and energy consumption results for all evaluated models under the two datasets, followed by a cross-dataset comparison.

\subsection{Accuracy–Energy Trade-off Analysis}
This subsection compares the selected DL architectures in terms of environmental impact, inference efficiency, and classification performance on the Breast Ultrasound and BreakHis datasets.

\subsubsection{Breast Ultrasound Dataset}
The inference performance of each model is detailed in Table~\ref{tab:accuracy}.\\
Results show that all models except DenseNet achieve over 80\% accuracy, a promising sign that the selected DL models perform well on this task regardless of architecture. Despite its high AUC, DenseNet achieved lower accuracy, indicating good class separability but more classification errors at the chosen decision threshold; this resulted in an overall less favourable performance profile than the competing models.
\begin{table}[!htbp]
\centering
\begin{tabular}{p{1.95cm}|p{0.85cm}|p{0.8cm}|p{1.2cm}|p{0.99cm}|p{0.55cm}}
\hline
\small
\textbf{Model} & Accuracy  & Recall* & F1-Score*& Precision*& AUC  \\ \hline
\textbf{ResNet50}      & 80.00\% & 86.66\%&86.98\% & 88.18\%&  96.71\% \\ 
\textbf{EfficientNetB4}  & 88.89\%& 84.76\% & 85.58\%& 87.90\%& 96.62\%\\  
\textbf{DenseNet121}& 76.67\% &74.28\% &89.51\% & 92.05\% & 97.65\%\\ 
\textbf{ViT-Tiny}          &   82.22\%& 85.71\% &  86.21\% & 87.31\%& 95.89\%\\          
\textbf{DeiT-Tiny}       & 86.67\% &92.38\%& 92.47\% &92.75\%&98.36\%\\             
\textbf{Swin-Tiny}          & 87.78\% &90.47\% & 90.49\%&90.58\% &98.35\%\\           
\textbf{ConvNeXt-Tiny}     &88.89\% &87.61\% & 87.82\% &88.62\%  &96.97\%\\ \hline
\end{tabular}
\caption{Inference Performance of the Selected DL Models on the Breast Ultrasound Dataset. *Weighted values.}
\label{tab:accuracy}
\end{table}
EfficientNetB4 and ConvNeXt-Tiny are the most performant models, both reaching 88.89\% accuracy. However, the result that stands out most belongs to DeiT-Tiny, despite not topping the accuracy, it leads every other model in recall (92.38\%), F1-score (92.48\%), precision (92.76\%) and ROC (98.36\%). In clinical practice, where an overlooked malignant case can have serious consequences, those numbers matter more than accuracy alone. Swin-Tiny also performs consistently well across all metrics. Taken together, transformer-based models bring a diagnostic reliability to this task that a single accuracy score simply cannot capture.

\begin{figure}[h]
    \centering
    \includegraphics[width=1\linewidth]{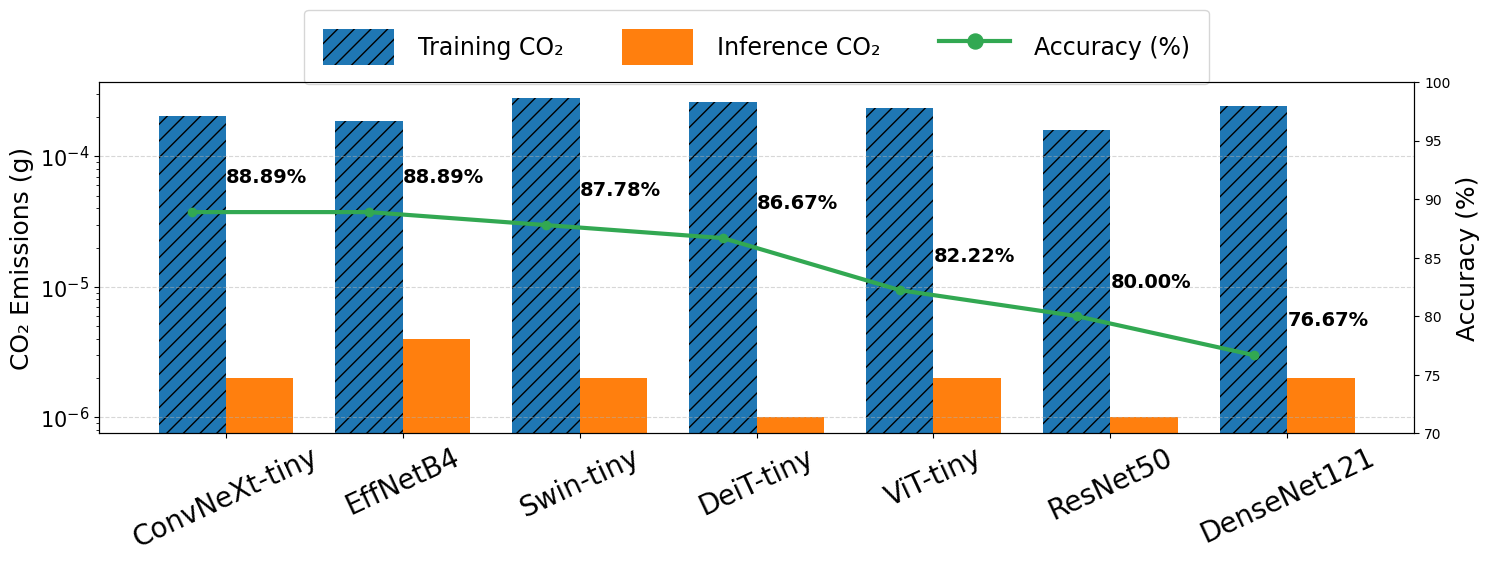}
    \caption{Training and Inference Emissions vs. Accuracy for the Selected Models on the Breast Ultrasound Dataset.}
    \label{fig:accuracyCarbonDataset1}
\end{figure}

To evaluate the environmental efficiency of the selected DL models, we quantify their carbon emissions during both training and inference phases and examine how these emissions relate to the models’ predictive accuracies. Figure~\ref{fig:accuracyCarbonDataset1} shows for each evaluated model the obtained accuracy and the CO$_2$ emissions consumed during the training and inference phases. \\ 

We observe that ResNet50 produces the lowest overall emissions, but its accuracy remains at 80.00\%, the second-lowest among all selected models. On the other hand, transformer models achieve up to 87.78\% accuracy but consume more energy. EfficientNetB4 and ConvNeXt-Tiny both achieve the highest accuracy of 88.89\%, although EfficientNetB4 produces the most inference emissions. Finally, ConvNeXt-Tiny achieves the same accuracy at a lower environmental cost, making it a more practical choice.

\begin{figure}[h]
    \centering
    \includegraphics[width=0.9\linewidth]{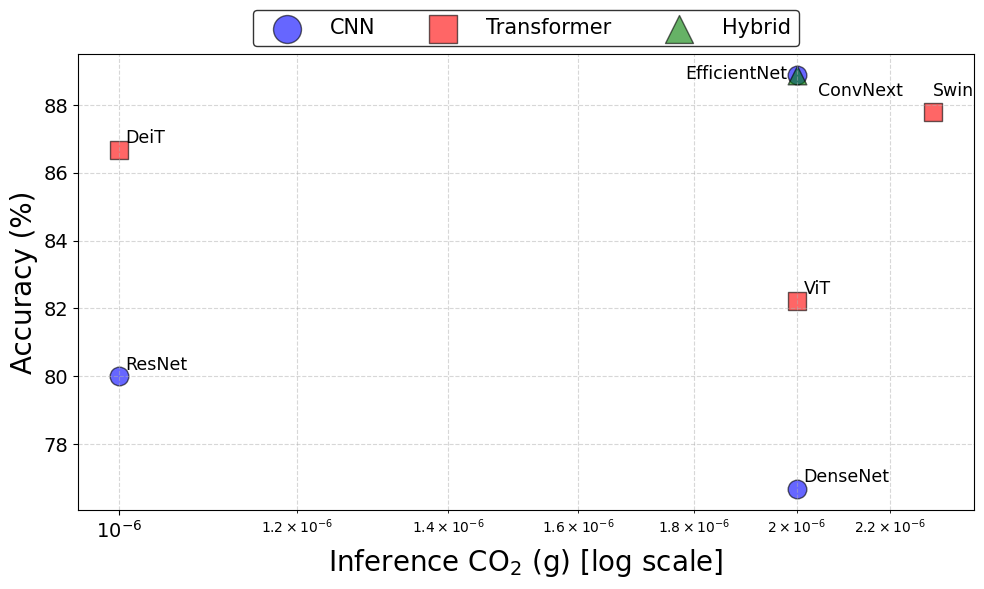}
    \caption{Classification Accuracy and Inference CO$_2$ Emissions (on a log scale) for the Selected Models on the Ultrasound Dataset.}
    \label{fig:inf_accuracy-dataset1}
\end{figure}

For real-world deployment, a detailed assessment of each model’s inference-related emissions and predictive performance is required. Figure~\ref{fig:inf_accuracy-dataset1} illustrates the relationship between these two metrics for the selected DL models.

According to Fig.~\ref{fig:inf_accuracy-dataset1}, ResNet and DeiT achieve the lowest inference emissions among all selected models, with DeiT achieving 86.67\% accuracy.
EfficientNet and ConvNeXt achieve the best accuracy 88.89\%, at the expense of high inference emissions. The Swin Transformer has slightly higher emissions without an accuracy gain. Moreover, DenseNet and ViT consume approximately as much energy for inference as EfficientNet and ConvNeXt, but with lower accuracy. 
In general, these findings demonstrate a trade-off between CO$_2$ emissions and accuracy. We conclude that higher emissions are not always correlated with better performance. DeiT-Tiny performs well, especially in terms of accuracy and recall, with low environmental cost, offering the best balance between performance and environmental impact. EfficientNetB4 and ConvNeXt-Tiny are the strongest models in terms of performance but consume more energy during inference. \\

\subsubsection{BreakHis Dataset} The inference performance metrics of the selected DL models in the BreakHis dataset are detailed in Table~\ref{tab:accuracy_models}.

\begin{table}[h]
\centering
\label{tab:accuracy_models}
\begin{tabular}{p{1.95cm}|p{0.9cm}|p{0.8cm}|p{1.2cm}|p{0.99cm}|p{0.65cm}}
\hline
\textbf{Model} & Accuracy  &Recall* &F1-Score*&Precision*&AUC  \\ \hline
\textbf{ResNet50 }      & 90.50\% &  90.49\%&  90.63\%&90.96\% & 96.16\%\\ 
\textbf{EfficientNetB4}  & 90.95\%& 90.95\% & 90.90\%&  90.88\% & 96.67\%\\  
\textbf{DenseNet121}& 74.21\% &72.85\%&72.14\% &  71.74\% & 75.84\%\\ 
\textbf{ViT-Tiny} &   89.14\%& 86.87\% & 87.28\%& 89.02\%& 94.99\% \\          
\textbf{DeiT-Tiny}       & 87.33\% & 83.25\% & 83.70\%&84.93\%&92.13\%\\             
\textbf{Swin-Tiny}          & 87.33\% &84.16\%& 84.44\%&85.01\%&91.92\%\\           
\textbf{ConvNeXt-Tiny} &86.43\% &86.42\% & 86.68\%& 87.32\%&92.27\%\\ \hline
\end{tabular}
\caption{Inference Performance of the Selected DL Models on the BreakHis Dataset. *Weighted values.}
\label{tab:accuracy_models}
\end{table}

As shown in Table~\ref{tab:accuracy_models}, EfficientNet and ResNet achieve the highest accuracies, around 90\%, supported by strong AUC values, making them the most competitive models across all performance metrics. ViT-Tiny follows closely behind, performing well and nearly matching top-evaluated CNN models. However, Swin-Tiny, ConvNeXt-Tiny, and DeiT perform lower on this dataset. Finally, DenseNet121 achieves the lowest accuracy among the evaluated models, at around 74.21\%. In general, from a performance perspective, EfficientNet and ResNet are the most reliable choices on this dataset, combining high accuracy with consistent robustness across evaluation metrics.

\begin{figure}[h]
    \centering
    \includegraphics[width=1\linewidth]{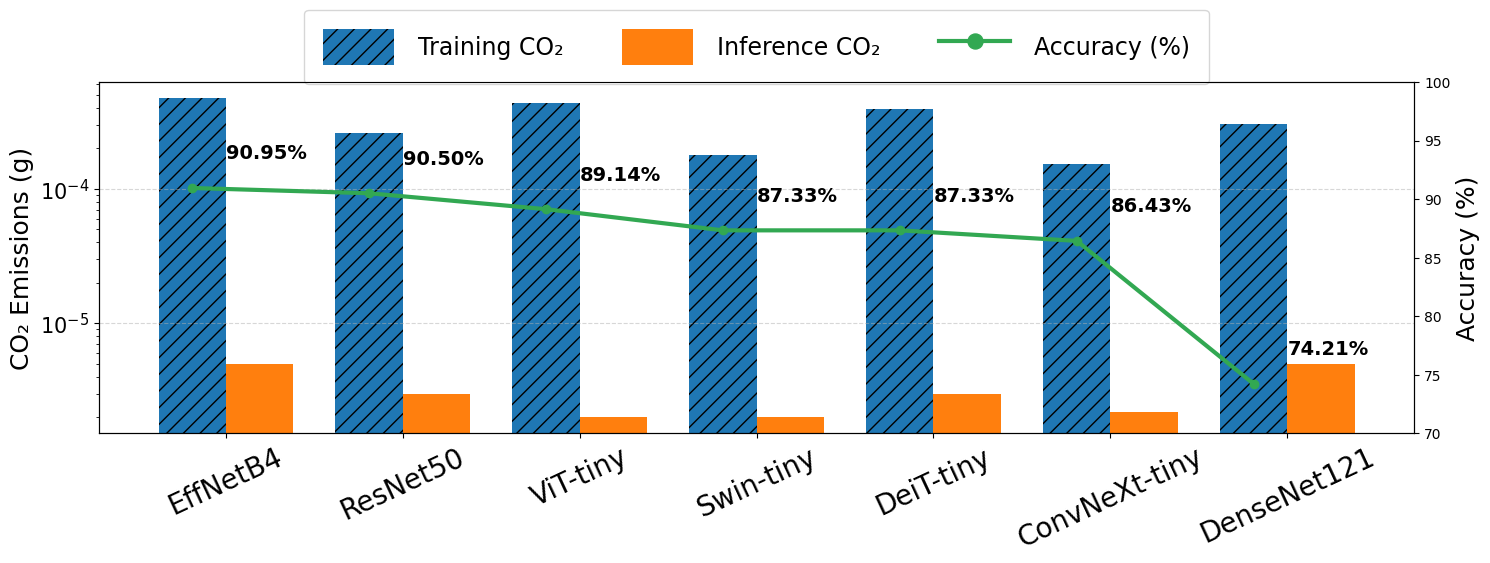}
    \caption{Training and Inference Emissions vs. Accuracy for the Selected Models on the BreakHis Dataset.¨}
    \label{fig:accuracy-carbon-dataset2}
\end{figure}

To assess environmental efficiency, Fig.~\ref{fig:accuracy-carbon-dataset2} shows classification accuracy and CO$_2$ emissions during both the training and inference phases for the selected DL models.

We observe a trade-off between predictive performance and environmental impact across the evaluated models. Swin-Tiny achieves a favourable balance by combining relatively high accuracy (~$\approx$87.33\%) with one of the lowest training and inference CO$_2$ emissions. EfficientNet-B4 and ResNet50 deliver the highest accuracies (~$\approx$90.95\% and ~$\approx$90.50\%, respectively), but at the cost of higher training-related carbon emissions, particularly for EfficientNet-B4. In contrast, ConvNeXt-Tiny exhibits low training CO$_2$ emissions but achieves a lower accuracy (~$\approx$86.43\%) compared with the best-performing models. These findings suggest that models such as Swin-Tiny can provide a more sustainable alternative by maintaining competitive predictive performance while reducing environmental impact.

\begin{figure}[h]
    \centering
    \includegraphics[width=0.9\linewidth]{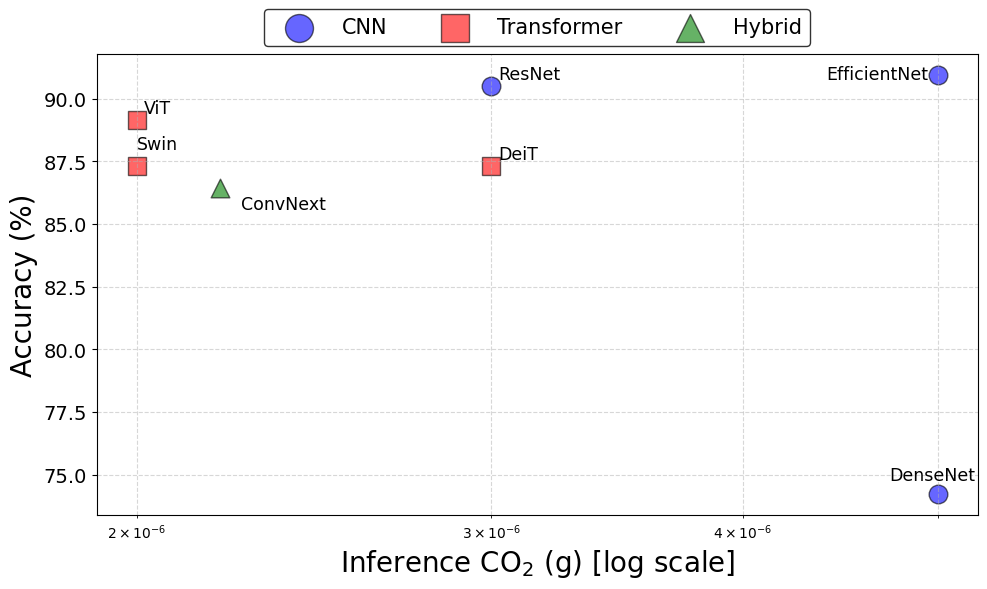}
    \caption{Classification Accuracy and Inference CO$_2$ Emissions (on a log scale) for the Selected Models on the BreakHis Dataset.}
    \label{fig:carbon-dataset2}
\end{figure}

To further investigate the relationship between accuracy and inference emission, a scatter diagram is provided in Fig.~\ref{fig:carbon-dataset2}. This diagram shows how inference CO$_2$ emissions (on a logarithmic scale) and classification accuracy relate to CNNs, transformers, and hybrid architectures.

We observe that CNNs, particularly EfficientNet and DenseNet, maintain exceptionally high inference CO$_2$ emissions. ResNet achieves the highest accuracy (90.5\%) among CNN models, with inference emissions much lower than EfficientNet. In addition, DenseNet shows the lowest accuracy (74\%) and the highest CO$_2$ emissions. On the other hand, ViT offers competitive accuracy (89.14\%) with a low carbon footprint, while Swin achieves moderate accuracy with similarly low emissions. ConvNeXt achieves moderate accuracy with low emissions. Overall, transformers, particularly ViT, provide the best trade-off between environmental efficiency and predictive performance during inference in the Breakhis dataset.

\subsection{Cross-Dataset comparaison}
With accuracies greater than 90\% on the BreakHis dataset and greater than 80\% on the Breast Ultrasound dataset, EfficientNetB4 and ResNet50 consistently rank among the best CNN-based models in both datasets. From an environmental standpoint, the results differ: EfficientNet exhibits among the highest CO$_2$ emissions during inference on both datasets, whereas ResNet’s inference emissions are low for the Breast Ultrasound dataset and moderate for the BreakHis dataset. On the other hand, DenseNet121 is the least appropriate architecture due to its continuously poor performance on both datasets, with the lowest accuracy values (76.67\% and 74.21\%) and the highest inference emissions. 

Regarding transformers, DeiT-Tiny performed well on both datasets with 87\% accuracy. Furthermore, it achieved the best recall (92.38\%) and ROC (98.36\%), and the lowest inference emissions, on the Breast Ultrasound Dataset. Finally, ConvNeXt-Tiny, the hybrid model, shows itself to be an attractive compromise, retaining competitive accuracy while minimizing carbon emissions.

Ultimately, the results indicate that no particular architecture, whether CNN-based, transformer-based, or hybrid, consistently dominates across datasets. High-performing models are found across all architectural categories, highlighting that performance is strongly influenced by the characteristics of the data set rather than the choice of the model alone.

\section{Threats to Validity}
\subsection{Internal Validity}
The data was split at the image level in the Breast Ultrasound dataset due to the absence of patient identifiers, which could introduce information leakage among the training, validation, and test sets. 
Energy consumption was estimated using the CodeCarbon estimation framework rather than direct power measurements. Consequently, the reported energy usage and associated carbon footprint values may be subject to estimation error.
\subsection{External Validity}
The small size of the Breast Ultrasound dataset may limit the generalizability of the results. For BreaKHis, only the classification head was trained while the pretrained backbone remained frozen.
\section{Conclusion and Future Work}
\label{sec:Conclusion}
This paper presents a comprehensive evaluation of seven DL models using two breast cancer imaging datasets, accounting for both performance and CO$_2$ emissions. 
Our main findings indicate that none of the evaluated architectures simultaneously achieves superior performance and energy efficiency across both datasets. EfficientNetB4 and ResNet50 consistently achieve the highest classification accuracy at comparatively higher environmental and energy costs. In contrast, transformer-based architectures such as DeiT-Tiny exhibit robust performance on both datasets, whereas DenseNet121 underperforms in all considered scenarios.  
For the Breast Ultrasound Dataset, the DeiT model offers the most favourable trade-off between accuracy and energy consumption. For the BreakHis dataset, the ViT and Swin architectures achieve the best energy-accuracy balance, ranking highest among the evaluated models.

More broadly, these findings show that accuracy alone should not drive model selection. In real-world applications, where resources are limited, both performance and computational efficiency must be considered. 

Although this study brings valuable findings, there are opportunities for future research. The results would benefit from multiple training runs to achieve more statistically robust conclusions, as well as the inclusion of another, larger, and more curated data set to improve generalization.

\phantomsection 
\addcontentsline{toc}{section}{References} 

\bibliographystyle{IEEEtran} 

\end{document}